\documentclass[pla,preprint,showpacs]{revtex4}
\usepackage{hyperref}
\usepackage{amsfonts}
\usepackage{amsmath}
\usepackage{amssymb}
\usepackage{graphicx}

\begin{document}

\title{Experimental creation of entanglement using separable state }
\author{Xiaodong Yang}
\author{An Min Wang}
\author{Xiaosan Ma}
\author{Feng Xu}
\author{Hao You}
\author{Wanqing Niu }
\affiliation{Department of Modern Physics, University of Science and Technology of China,
Hefei, 230026, People's Republic of China}
\pacs{03.67.Mn, 76.60.-k}

\begin{abstract}
We experimentally demonstrate the entanglement can be created on two distant
particles using separate state. We show that two data particles can share
some entanglement while one ancilla particle always remains separable from
them during the experimental evolution of the system. Our experiment can be
viewed as a benchmark to illustrate the idea that no prior entanglement is
necessary to create entanglement.
\end{abstract}

\maketitle

\section{Introduction}

Entanglement is one of the inherent features of quantum mechanics \cite{epr}%
. It is unambiguously a crucial resource and also plays an important role in
most of quantum information processing (QIP) tasks, e.g., \cite%
{bennett1,bennett2}. It becomes well worth investigating how entanglement is
available.

There have been several proposals to create entanglement. (i) Applying a
global quantum operation on the interested system \cite{barenco}. For
simplicity, we consider the initial state is the ground state $\left\vert
00\ldots 0\right\rangle $ in an $N$-qubit system. The entangled state $%
(\left\vert 00\ldots 0\right\rangle +\left\vert 11\ldots 1\right\rangle )/%
\sqrt{2}$ can be created by a Hadamard gate followed by $(N-1)$
controlled-not (C{\small NOT}) gates. Based on this idea, the three-qubit
Greenberger-Horne-Zeilinger (GHZ) state was generated experimentally \cite%
{lafla}. Recently, Zhou \textit{et al. }\cite{zhou} proposed an evolution
method to create multi-qubit maximally entangled state in the nuclear
magnetic resonance (NMR) model, and their method need only a single-step
operation. Note that such kind of methods requires the interaction between
the particles. However, this direct interaction is not always existing in a
real physical system. Therefore, we should consider the situation under the
separated condition. (ii) Making use of the swap operation \cite{madi}. The
basic idea is that firstly entangling two ancilla particles which are
directly interacted, and then transferring this entanglement to another pair
of data particles which are separable via two swap gates. This method, more
precisely speaking, can be called as \textit{entanglement transfer}.
Recently Boulant \textit{et al.} \cite{boulant1} have demonstrated this idea
on a four-qubit liquid NMR quantum processor. (iii) Distributing
entanglement using a separable state. This idea was proposed by Cubbit
\textit{et al.} \cite{cubitt}. They show that no prior entanglement is
necessary to create entanglement.

A few of comments about the differences between the method (iii) and the
former two are in order. First, the method (iii) can be used in the
situation where there are no direct interactions among the particles, which
is the difference from the method (i). Second, the swap operation is not
used in the method (iii), which is the difference from the method (ii). Last
but not the least, the method (iii) clarifies the fact that two separable
particles can be entangled without the ancilla particle ever becoming
entangled with them.

In this Letter, we report an experimental creation of entanglement based on
the third method mentioned above using NMR technique. The problem of
interest in this Letter is the experimental demonstration of entanglement
creation between two non-directly-coupled spins with the aid of one ancilla
spin, and most importantly, keeping the ancilla never entangled with the
data spins. Our results confirm the separable state can be used to create
entanglement.

\section{Theory of entanglement creation using separable state}

We begin with a three-particle system --- A, B and C, each of which is a
spin-half quantum bit (qubit). Suppose the two data qubits A and B are so
well isolated or far away over a long distance that they do not directly
interact with each other, while ancilla C interacts with both A and B. Our
goal is to entangle A and B, or at least entangle them with some
probability. An explicit case \cite{cubitt} is given as follows.

The initial state $\rho _{0}$ is well chosen to satisfy the condition that
the system ABC is separable at the first beginning and can finally lead to
the entangled state in AB system after some certain quantum operations. We
choose $\rho _{0}$ as a mixed state composed of the six pure ones (note that
the pure state in ABC is impossible to achieve our goal \cite{cubitt}),

\begin{align}
\rho _{0}& =\frac{1}{6}P[\frac{1}{\sqrt{2}}\left( \left\vert 0\right\rangle
+\left\vert 1\right\rangle \right) \otimes \frac{1}{\sqrt{2}}\left(
\left\vert 0\right\rangle +\left\vert 1\right\rangle \right) \otimes
\left\vert 0\right\rangle ]  \notag \\
& +\frac{1}{6}P[\frac{1}{\sqrt{2}}\left( \left\vert 0\right\rangle
+i\left\vert 1\right\rangle \right) \otimes \frac{1}{\sqrt{2}}\left(
\left\vert 0\right\rangle -i\left\vert 1\right\rangle \right) \otimes
\left\vert 0\right\rangle ]  \notag \\
& +\frac{1}{6}P[\frac{1}{\sqrt{2}}\left( \left\vert 0\right\rangle
-\left\vert 1\right\rangle \right) \otimes \frac{1}{\sqrt{2}}\left(
\left\vert 0\right\rangle -\left\vert 1\right\rangle \right) \otimes
\left\vert 0\right\rangle ]  \notag \\
& +\frac{1}{6}P[\frac{1}{\sqrt{2}}\left( \left\vert 0\right\rangle
-i\left\vert 1\right\rangle \right) \otimes \frac{1}{\sqrt{2}}\left(
\left\vert 0\right\rangle +i\left\vert 1\right\rangle \right) \otimes
\left\vert 0\right\rangle ]  \notag \\
& +\frac{1}{6}P[\left\vert 001\right\rangle ]+\frac{1}{6}P[\left\vert
111\right\rangle ],  \label{ini}
\end{align}%
where the symbol $P[\left\vert \varphi \right\rangle ]=\left\vert \varphi
\right\rangle \left\langle \varphi \right\vert $ denotes the density matrix
corresponding to state $\left\vert \varphi \right\rangle $. From Eq. (\ref%
{ini}), A, B and C are separable from each other.

Next, apply a C{\small NOT} gate on qubits A and C, where A is the control
qubit and C the target qubit. From nontrivial calculation, the initial state
(\ref{ini}) will become as

\bigskip
\begin{align}
\rho _{1}& =\frac{1}{3}P[\left\vert \Phi _{GHZ}\right\rangle ]+\frac{1}{6}%
P[\left\vert 001\right\rangle ]  \notag \\
& +\frac{1}{6}P[\left\vert 010\right\rangle ]+\frac{1}{6}P[\left\vert
101\right\rangle ]+\frac{1}{6}P[\left\vert 110\right\rangle ],  \label{rou1}
\end{align}%
where $\left\vert \Phi _{GHZ}\right\rangle =\frac{1}{\sqrt{2}}\left(
\left\vert 000\right\rangle +\left\vert 111\right\rangle \right) $ is the
GHZ state. Note the difference of the phases in the singlet states in the
initial state (\ref{ini}) is counteracted after the C{\small NOT} operation.
In this state, we can check \cite{cubitt} that C is separable from AB, B is
separable from AC, and only A is entangled with BC.

Later, send the qubit C to B through quantum channel, and then apply another
C{\small NOT} gate on qubits B and C, where B is the control qubit and C the
target qubit. Thus, the final state will be

\begin{equation}
\rho _{2}=\frac{1}{3}P[\left\vert \phi ^{+}\right\rangle _{AB}\otimes
\left\vert 0\right\rangle _{C}]+\frac{2}{3}(\frac{1}{4}I)_{AB}\otimes
\left\vert 1\right\rangle _{C}\left\langle 1\right\vert ],  \label{fin}
\end{equation}%
where $\left\vert \phi ^{+}\right\rangle =\frac{1}{\sqrt{2}}\left(
\left\vert 00\right\rangle +\left\vert 11\right\rangle \right) $ is the Bell
state, and $I$ is the unit matrix of order 4. From Eq. (\ref{fin}), C still
remains separable from AB, but the entanglement between A and B is produced.
Obviously, measuring C in the computational basis, the entangled state $%
\left\vert \phi ^{+}\right\rangle $ of AB will be extracted with the
probability 1/3. Note that the AB system after tracing out C is not an
entangled state, because after tracing, AB is in a type of Werner state \cite%
{werner} $\frac{1}{3}\left\vert \phi ^{+}\right\rangle \left\langle \phi
^{+}\right\vert +\frac{2}{3}\cdot \frac{I}{4}$, which is just separable.

It is well known that entanglement can be created on two distant qubits by
sending a mediating (ancilla) particle between them, and many multi-qubit
quantum information transmission protocols are based on this kind of
entanglement creation \cite{luo,wei}. One may expect that the ancilla
necessarily becomes entangled with the system. However, from the above
analysis, the scheme proved that the ancilla C can never entangled with the
two data qubits AB although we used the interactions between C and AB.

\section{Experimental demonstration}

NMR is one of the most important physical system to explore the
implementation of QIP experimentally, especially in the few-qubit system.
NMR quantum processor has been widely used to test many kinds of QIP tasks
(for a review see e.g., \cite{jones, cory1}). The nature of NMR quantum
computing is reinvestigated recently \cite{long}.

\begin{figure}[tbp]
\includegraphics[scale=1]{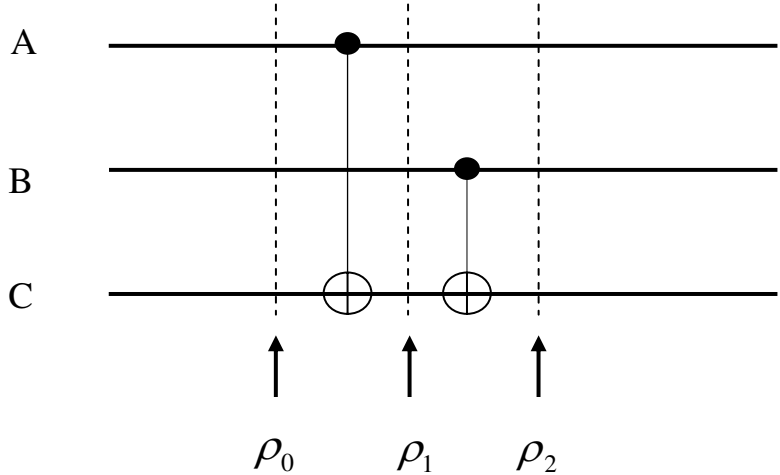}
\caption{Quantum circuit to create entanglement between two distant qubits A
and B, while qubit C works as an ancilla, which is never entangled with AB
during our scheme. Time flows from left to right. $\protect\rho _{0}$ is the
initial state, $\protect\rho _{1}$ is the state after the CNOT gate on AC,
and $\protect\rho _{2}$ is the final state after the CNOT gate on BC. For
the states $\protect\rho _{0}$, $\protect\rho _{1}$, $\protect\rho _{2}$, we
reconstruct their density matrices.}
\label{mypic1}
\end{figure}

Since the nuclear spins are fixed in the molecule through chemical bonds in
NMR experiment, there is a distance of only a few angstrom ($\mathring{A}$)
between different spins, thus it is difficult to realize quantum channel in
nuclear spin system \cite{nielsen}. Our experiment made a demonstration of
quantum communication, rather than a practical means for sending information
through quantum channel on distant particles. The whole demonstration
procedure is shown in Fig. \ref{mypic1}.

We performed the experiment on a Bruker AVANCE 400 MHz spectrometer, keeping
temperature at 300K. The spin system is $^{13}$C-labeled alanine $%
NH_{3}^{+}-C^{\alpha }H(C^{\beta }H_{3})-C^{\prime }OO^{-}$ \cite{du}, where
carbons $C^{\prime },C^{\beta },C^{\alpha }$ correspond to qubits A, B and
C, respectively. In the following, we label the three qubits A, B, C as 1,
2, 3 in keeping with the conventional parlance. The coupling constants
between three carbons\ are

\begin{equation*}
J_{13}=54.2Hz,J_{23}=35.1Hz,J_{12}=-1.3Hz.
\end{equation*}%
In our experiment, $J_{12}$ is always refocused though the coupling constant
is small. This condition means that qubits 1 and 2 have no interaction which
agrees with our demand. The experimental process is explained as the
following three stages:

$(1)$ Prepare for the initial state as shown in Eq. (\ref{ini}). We firstly
assemble the pseudopure state using the gradient-based spatial averaging
\cite{cory2}. Note that since the NMR system is a spin ensemble, it is not
easy to prepare for a pure state experimentally. However, the unitary
dynamics of the pseudopure state is same as the pure state up to a certain
scaling factor \cite{gershenfeld,cory3}. The pseudopure state, keeping the
traceless part, can be expressed with the product operators formalism \cite%
{ernst} which has been commonly used in NMR community,

\begin{align}
\rho _{pp}& =\left\vert 000\right\rangle \left\langle 000\right\vert
\label{pure} \\
&
=I_{z}^{1}+I_{z}^{2}+I_{z}^{3}+2I_{z}^{1}I_{z}^{2}+2I_{z}^{1}I_{z}^{3}+2I_{z}^{2}I_{z}^{3}+4I_{z}^{1}I_{z}^{2}I_{z}^{3},
\notag
\end{align}%
where $I_{\alpha }^{i}=\frac{1}{2}\sigma _{\alpha }^{i}$ $(i=1,2,3,\alpha
=x,y,z)$, $\sigma _{\alpha }^{{}}$ is the Pauli matrix.

In order to obtain the initial state from the pseudopure state, we note that
the initial state (\ref{ini}) is a six-componential mixed state. Our scheme
involves six experiments and a sum of these six results. The scheme to
prepare for the initial state is shown in table \ref{table1}.

\begin{table}[tbp]
\caption{The six components in the initial state and the corresponding pulse
sequences for the preparation of the initial state. In the right column, $(%
\protect\theta )_{axis}^{spins}$ denote the radio frequency pulses, which
are applied to the spins in the superscript, along the axis in the
subscript, and by the flip angle in the bracket.}
\label{table1}%
\begin{ruledtabular}
\begin{tabular}{cc}
\multicolumn{1}{c}{Components in the initial state}&\multicolumn{1}{c}{Pulse sequence}\\
\hline

 $\frac{1}{\sqrt{2}}\left( \left\vert 0\right\rangle +\left\vert
1\right\rangle \right) \otimes\frac{1}{\sqrt{2}}\left( \left\vert
0\right\rangle +\left\vert 1\right\rangle \right)
\otimes\left\vert
0\right\rangle $&$\left( \frac{\pi}{2}\right) _{y}^{1,2}$ \\

$\frac {1}{\sqrt{2}}\left( \left\vert 0\right\rangle +i\left\vert
1\right\rangle \right) \otimes\frac{1}{\sqrt{2}}\left( \left\vert
0\right\rangle -i\left\vert 1\right\rangle \right)
\otimes\left\vert 0\right\rangle $&$\left( \frac{\pi}{2}\right)
_{-x}^{1}\left( \frac{\pi}{2}\right) _{x}^{2}$\\

$\frac{1}{\sqrt{2}}\left( \left\vert 0\right\rangle -\left\vert
1\right\rangle \right) \otimes \frac{1}{\sqrt{2}}\left( \left\vert
0\right\rangle -\left\vert 1\right\rangle \right) \otimes
\left\vert 0\right\rangle $&$\left( \frac{\pi}{2}\right)
_{-y}^{1,2}$\\

$\frac{1}{\sqrt{2}}\left( \left\vert 0\right\rangle -i\left\vert
1\right\rangle \right) \otimes \frac{1}{\sqrt{2}}\left( \left\vert
0\right\rangle +i\left\vert 1\right\rangle \right) \otimes
\left\vert 0\right\rangle $&$\left( \frac{\pi}{2}\right)
_{x}^{1}\left( \frac{\pi}{2}\right)
_{-x}^{2}$\\

$\left\vert 001\right\rangle$&$\left( \pi\right) _{y}^{3}$\\

$\left\vert 111\right\rangle $&$\left( \pi\right) _{y}^{1,2,3}$\\
\end{tabular}
\end{ruledtabular}
\end{table}
The summed result corresponds to the initial state (\ref{ini}) , whose
expression in terms of the product operators is

\begin{equation}
\rho_{0}=I_{z}^{3}+2I_{x}^{1}I_{x}^{2}-2I_{y}^{1}I_{y}^{2}+2I_{z}^{1}I_{z}^{2}+4I_{x}^{1}I_{x}^{2}I_{z}^{3}-4I_{y}^{1}I_{y}^{2}I_{z}^{3}-4I_{z}^{1}I_{z}^{2}I_{z}^{3}.
\end{equation}

$(2)$ Apply a C{\small NOT} operation on AC. The pulse sequence for
realizing the C{\small NOT } gate \cite{jones,vander} is $\left( \frac{\pi }{%
2}\right) _{y}^{3}-\frac{1}{2J_{13}}-\left( \frac{\pi }{2}\right)
_{x}^{3}-\left( \frac{\pi }{2}\right) _{-z}^{3}-\left( \frac{\pi }{2}\right)
_{z}^{1}$, where the symbol $\frac{1}{2J_{ij}}$ denotes the evolution time
dominated by the J coupling between spins $i$\ and $j$; $z$ pulse on spin $i$
is implemented using the combination of $x$ and $y$ pulses $\left( \frac{\pi
}{2}\right) _{x}^{i}-\left( \frac{\pi }{2}\right) _{y}^{i}-\left( \frac{\pi
}{2}\right) _{-x}^{i}$. During the evolution time $\frac{1}{2J_{13}}$, the
refocusing pulses on spins 1 and 3 are applied to eliminate not only the
coupling between B and AC, but also the effect of the chemical shift. The
state after this stage is

\begin{equation}
\rho_{1}=2I_{z}^{1}I_{z}^{2}+2I_{z}^{1}I_{z}^{3}-2I_{z}^{2}I_{z}^{3}-4I_{y}^{1}I_{x}^{2}I_{y}^{3}-4I_{x}^{1}I_{y}^{2}I_{y}^{3}+4I_{x}^{1}I_{x}^{2}I_{x}^{3}-4I_{y}^{1}I_{y}^{2}I_{x}^{3}.
\end{equation}

$(3)$ Apply another C{\small NOT} operation on BC and obtain the final
state, which can be expressed as

\begin{equation}
\rho
_{2}=-I_{z}^{3}+2I_{x}^{1}I_{x}^{2}-2I_{y}^{1}I_{y}^{2}+2I_{z}^{1}I_{z}^{2}+4I_{x}^{1}I_{x}^{2}I_{z}^{3}-4I_{y}^{1}I_{y}^{2}I_{z}^{3}+4I_{z}^{1}I_{z}^{2}I_{z}^{3},
\end{equation}%
where the final state includes the entanglement of AB. If performing a
projective measurement on C, we can extract the entanglement on AB with
probability $1/3$. However, NMR measure is a spatial ensemble average or an
expected value \cite{cory1}, rather than a projective measurement. Though we
can mimic the projective (strong) measurement by magnetic field gradients
\cite{tekle1} or natural decoherence \cite{nielsen}, these two techniques do
not adapted here because both of them can not distinguish whether the
measured qubit is projected into space $\left\vert 0\right\rangle
\left\langle 0\right\vert $ or $\left\vert 1\right\rangle \left\langle
1\right\vert $. In other words, after the mimic projective measurement, the
state is that of tracing out the measured qubit, but this state, as
mentioned above, is not an entangled state.

\begin{figure}[tbp]
\begin{center}
\includegraphics[scale=1]{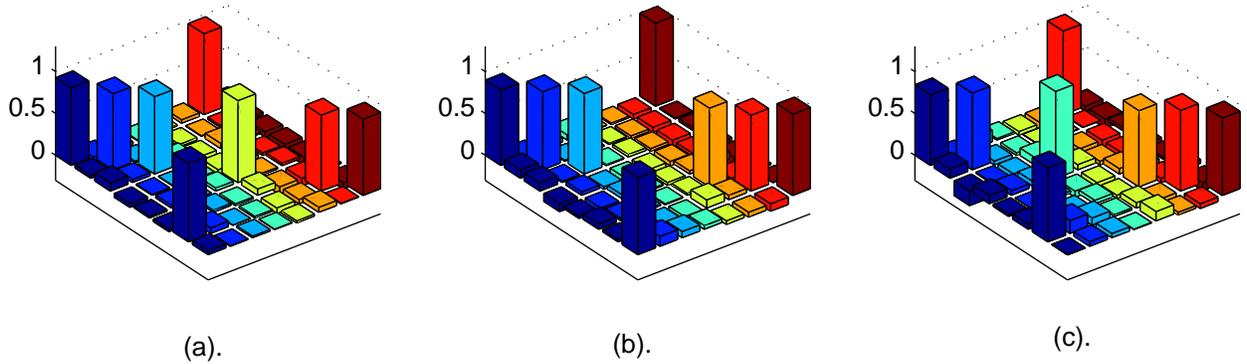}
\end{center}
\caption{Real parts of the reconstructed density matrices of (a).the initial
state $\protect\rho_{0}$, (b).the medial state $\protect\rho_{1}$, and
(c).the final state $\protect\rho_{2}$. The imaginary parts are essentially
zero. The rows and columns represent the standard computational basis, with $%
\left\vert 000\right\rangle $ starting from the leftmost column and $%
\left\vert 111\right\rangle $ the rightmost column.}
\label{mypic2}
\end{figure}

As a consequence, to confirm the entanglement existing in the final state,
we use the state tomography technique \cite{chuang} to reconstruct the
density matrices at the corresponding three experimental stages. The results
are shown in Fig. \ref{mypic2}.

The experimental efficacy was quantified by the \textit{attenuated
correlation} \cite{tekle1}, which takes into account not only systematic
errors in experiment, but also the random errors. This measure is given by

\begin{equation}
C(\rho _{{}}^{exp})=\frac{Tr\left( \rho _{{}}^{exp}\rho _{{}}^{the}\right) }{%
Tr\left( \rho _{{}}^{the}\rho _{{}}^{the}\right) }
\end{equation}%
where $\rho _{{}}^{the}$ is the measured pseudopure ground state
(the reconstructed density matrix is not shown here); $\rho
_{{}}^{exp}$ represents the experimental state realized by a
series of pulse sequences. The values of the correlation for the
three states $\rho _{0},\rho _{1},\rho _{2}$ are $C(\rho
_{0}^{exp})=98\%,C(\rho _{1}^{exp})=95\%,C(\rho _{2}^{exp})=89\%$,
respectively. The correlation values show that spins 1
and 2 are separable in both $\rho _{0}$ and $\rho _{1}$, but entangled in $%
\rho _{2}$, while spin 3 keeps separable in $\rho _{0},\rho _{1},\rho _{2}$
by the corresponding to Eqs. (\ref{ini}), (\ref{rou1}), (\ref{fin}). The
loss of correlation mainly includes the imperfect selective pulses, and the
variability during the measurement process.

\section{Conclusion}

We demonstrated that entanglement can be produced using separated state with
NMR technique. In our scheme, there is only one ancilla qubit required to
obtain the entanglement on two qubits which have no direct interaction.
Compared with the entanglement swapping \cite{boulant1} where two ancilla
qubits are required, the number of the ancilla is reduced. This suggests
that our experimental method is less demanding on the qubit resource. On the
other hand, we show that if we select a proper initial state, the ancilla
qubit will never entangle with the data qubits during the evolution of the
system, which is completely different form the common used method to create
entanglement on distant qubits \cite{wei}. This also shows a striking fact
that no prior entanglement is required to create entanglement. Moreover,
this scheme can be extended to multi-qubit system where there are no direct
interactions among the qubits \cite{cubitt}.

It should be noted that, using this method to create entanglement, a
projective measurement is needed to extract the $(n-1)$-qubit entanglement
from the $n$-qubit system for the aim of later QIP task. Alternatively, one
also can extract the useful information from the final state after the whole
QIP task according to the principle of state superposition and parallelism.

\begin{acknowledgements}
We thank Jiang-Feng Du, Ming-Jun Shi, and Ping Zou for useful
discussions. This work was supported by China Post-doctoral
Science Foundation, the National Basic Research Programme of China
under Grant No 2001CB309310, the National Natural Science
Foundation of China under Grant No 60173047, and the Natural
Science Foundation of Anhui Province.
\end{acknowledgements}

\end{document}